%% file: paper.tex
\documentclass[graybox,natbib,smallextended]{svmult}

\usepackage{amsmath}%
\usepackage{amssymb}%
\usepackage{wasysym}%

\usepackage{mathptmx}       
\usepackage{helvet}             
\usepackage{courier}           
\usepackage{type1cm}         
\usepackage{makeidx}              
\usepackage{graphicx}              
\usepackage{multicol}                
\usepackage[bottom]{footmisc}  

\makeindex             

\usepackage{color}
\usepackage{hyperref}
\usepackage{url}
\usepackage{soul} 
\usepackage{ulem} 

\input{localdefs.tex}

\begin{document}

\title*{Connecting Numerical Relativity and Data Analysis of Gravitational Wave Detectors}
\author{Deirdre Shoemaker, Karan Jani, Lionel London, Larne Pekowsky}
\institute{Deirdre Shoeamer \at Center for Relativistic Astrophysics, School of Physics, Georgia Institute of Technology, Atlanta GA 30332 \email{deirdre@gatech.edu}
\and Karan Jani \at Center for Relativistic Astrophysics, School of Physics, Georgia Institute of Technology, Atlanta GA 30332 \email{kjani3@gatech.edu}
\and Lionel Londoni \at Center for Relativistic Astrophysics, School of Physics, Georgia Institute of Technology, Atlanta GA 30332 \email{london@gatech.edu}
\and Larne Pekowsky \at Center for Relativistic Astrophysics, School of Physics, Georgia Institute of Technology, Atlanta GA 30332 \email{larne.pekowsky@physics.gatech.edu}
}

\maketitle

\abstract{Gravitational waves deliver information in exquisite detail about astrophysical phenomena,
among them the collision of two black holes, a system completely invisible to the eyes of electromagnetic
telescopes.  Models that predict gravitational wave signals from likely sources are crucial
for the success of this endeavor. Modeling binary black hole sources of gravitational radiation requires solving the
Eintein equations of General Relativity using powerful computer hardware and sophisticated
numerical algorithms.  This proceeding presents where we are in understanding ground-based gravitational waves resulting from the merger of black holes and the implications of these sources for the advent of gravitational-wave astronomy.}

\section{Introduction}
\label{intro}
During the GR 1 conference  held on January 1957 at the University of North Carolina, Chapel Hill,  Charlie Misner famously said, ``... we assume that we have a computer machine better than anything we have now, and many programmers and a lot of money, and you want to look at a nice pretty solution of the Einstein equations. ... Now, if you don't watch  out when you specify these initial conditions, then either the programmer will shoot himself or the machine will blow up~\citep{GR1}."  From that date in 1957 to Frans Pretorius' landmark paper in 2005~\citep{Pretorius:2005gq}  that successfully followed an orbit and merger of two \bh{s}, many programmers and money tackled the problem that Misner so eloquently points out. With two other groups following the Pretorius paper within a year~\citep{Baker:2005vv,Campanelli:2005dd}, a new age of \nr{} was begun.  Progress was being made  over the intervening decades~\citep{Brandt:1997tf,Bruegmann:1997uc,Abrahams:1997ut,Cook:1997na,Gomez:1998uj,Brandt:2000yp,Alcubierre:2000ke,Alcubierre:2002kk,Bruegmann:2003aw,Gundlach:2005eh} to reference just a few of the many pivotal papers during that time.

Arguably the driving force behind the decades of work on solving Einstein's equations for the coalescence of compact binary systems has been and continues to be the imminent detection of \gw{s}.  The world-wide network of detectors is coming online with  Advanced LIGO~\citep{Harry:2010zz} and Advanced Virgo~\citep{Acernese:2009} leading the way and KAGRA~\citep{Somiya:2011np} on the way.  The era of gravitational wave astronomy is upon us~\citep{Aasi:2013wya} and NR has an important role to play.   Just some of the important issues currently being addressed by NR for GW astronomy are
\begin{itemize}
\item determining where approximations to full solutions to the Einstein equation break down as we approach merger~\citep{Ohme:2011zm,MacDonald:2012mp},
\item detecting the difference between NS and BH sources~\citep{Foucart:2013psa,Hannam:2013uu},
\item creating template banks that cover the complete parameter space~\citep{Kumar:2013gwa,Privitera:2013xza},
\item determining the impact of higher harmonics on detection and parameter estimation,
\item settling down to final state of the black hole, and
\item investigating mergers as GW transients.
\end{itemize}

This proceeding  focuses on that last three items in the above list.  First, we discuss a historical perspective of NR and GWs in $\S$\ref{hist} and a more detailed look into BBH mergers in NR $\S$\ref{bbh} before discussing the role of harmonics $\S$\ref{modes}, the approach to the final state $\S$\ref{endstate} and mergers as transients $\S$\ref{bursts}.  Conclusions are presented in $\S$\ref{concl}.

\section{Historical Perspective on NR and GWs}
\label{hist}

While NR was struggling to capture the spacetime and dynamics of a merging compact binary on computers, the gravitational wave community was preparing for initial LIGO and VIRGO.  Ref.~\citep{Cutler:1992tc} lays out the case for a deep understanding of the theoretical waveforms, called templates, from compact binary coalescence in preparation for the GW detectors.  These theoretical templates are needed to unearth the signal within the noisy data using matched filtering, the optimal choice for detecting a signal in Gaussian noise where the template is known~\citep{Thorne}.  Although at the time NR had not yet produced waveforms from BBH mergers, post-Newtonian theory was making progress in producing waveforms that work well for binary neutron stars and low mass \bh{} binaries.  Low-mass objects, a couple of solar masses to tens of solar masses, have long inspirals in the LIGO frequency band well modeled by post-Newtonian approximations. In fact, neutron star binaries may merge outside of the band making NR less useful in predicting these signals~\footnote{ To understand matter's effect on neutron star waveforms see Ref~\citep{Read:2013zra} for binary neutron stars and Ref.~\citep{Andersson:1997rn} for isolated neutron stars. }

BBH signals that are more massive could be the first detected~\citep{Brady:1998du} motivating several methods to bridge the gap between post-Newtonian inspiral and ringdown.  Energy conditions were studied \citep{Flanagan:1997sx,Flanagan:1997kp} and approximations were developed to produce waveforms that would cover the full inspiral, merger, ringdown \citep{Buonanno:1998gg,Khanna:1999mh,Baker:2001sf}.  Fortunately, NR was on the brink of success culminating in the series of papers in 2005-2006~\citep{Pretorius:2005gq,Campanelli:2005dd,Baker:2005vv}.

\section{Binary black hole mergers}
\label{bbh}
Post 2005, NR groups world wide began an assault on the BBH parameter space, focusing primarily on gravitational recoil and the spin of the final black hole.  A few turned their attention to the ramifications of these new NR waveforms on data analysis efforts to detect BBHs~\citep{Baumgarte:2006en,Buonanno:2006ui,Baker:2006kr,Ajith:2007qp,Pan:2007nw,Vaishnav:2007nm}.   This culminated in three collaborations
\begin{enumerate}
\item NINJA~\citep{Aylott:2009ya,Aylott:2009tn,Ajith:2012az,Aasi:2014tra}: the Numerical INJection Analysis project whose goal is to use NR waveforms as signal and test the detection and parameter estimation pipelines for the ground-based GW detectors,
\item SAMURI \citep{Hannam:2009hh}: this was a one-time project that did a consistency check of several NR codes; and,
\item NRAR~\citep{Hinder:2013oqa}: Numerical Relativity and Analytical Relativity Collaboration whose goal is to use NR waveforms to produce the next generation of waveform models.
\end{enumerate}
For more details on these collaborative efforts in the numerical relativity and gravitational wave community, see the paper by Sascha Husa in this collection.

\begin{figure}[h]
\centerline{
\includegraphics[scale=.5,trim= 50 150 0 150,clip]{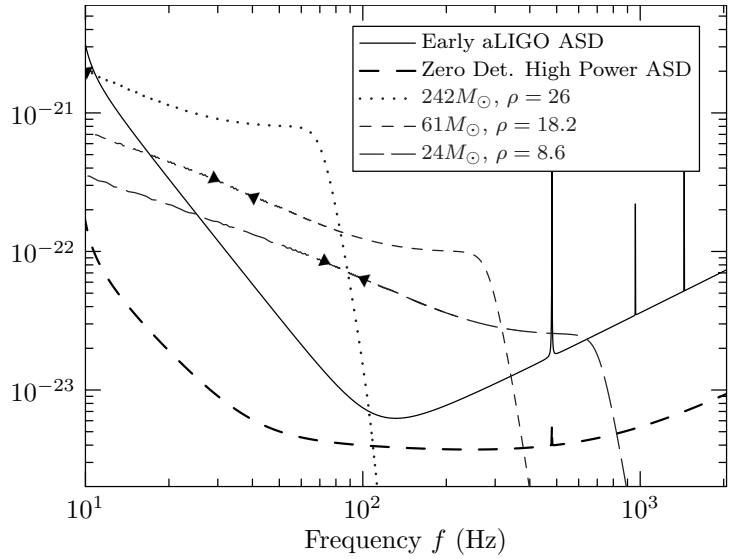}
}
\caption{A Ninja 2 Collaboration~\citep{Ajith:2012az} figure.  Here we see the GW from a NINJA2 waveform versus frequency against two predictions of the advanced LIGO noise curve.  A single NR waveform is represented at three different total masses.  The waveform is a hybrid, non-spinning, 2 to 1 mass ratio waveform scaled to three different total masses.  The triangles represent the starting and ending frequencies of the hybridization region where the NR and post-Newtonian waveforms were stitched together.  The waveform is scaled such that it is optimally oriented at a distance of 1 Gpc from the detector.  The signal-to-noise ratios, $\rho$, were computed at that distance with the early aLIGO sensitivity.}
\label{f:ninja2}       
\end{figure}

Figure~\ref{f:ninja2} is a figure from the NINJA 2 paper~\citep{Ajith:2012az} where a catalog of NINJA waveforms were publicly  released. The figure depicts one hybrid waveform at three different total masses.   The merger is at the extremum of the waveform.  The triangles represent the region where we created hybrids by stitching together the post-Newtonian  NR waveforms.  As mass increases, the signal-to-noise ratio increases and the number of cycles in band decreases at a fixed distance for an optimally oriented binary with the parameters held fixed; however, at some point, the reduction in the number of cycles will result in a drop of signal-to-noise ratio.
   The figure illustrates the impact of merger on the BBH signals for masses above $20 M_\odot$ where the merger sits in the most sensitive region of the detector.  As the masses grow higher than a couple of hundred solar masses, we begin to only detect the merger and ringdown addressed in $\S$\ref{bursts}.  

\begin{figure}[h]
\hbox{
\includegraphics[scale=.3,trim= 450 100 300 100,clip]{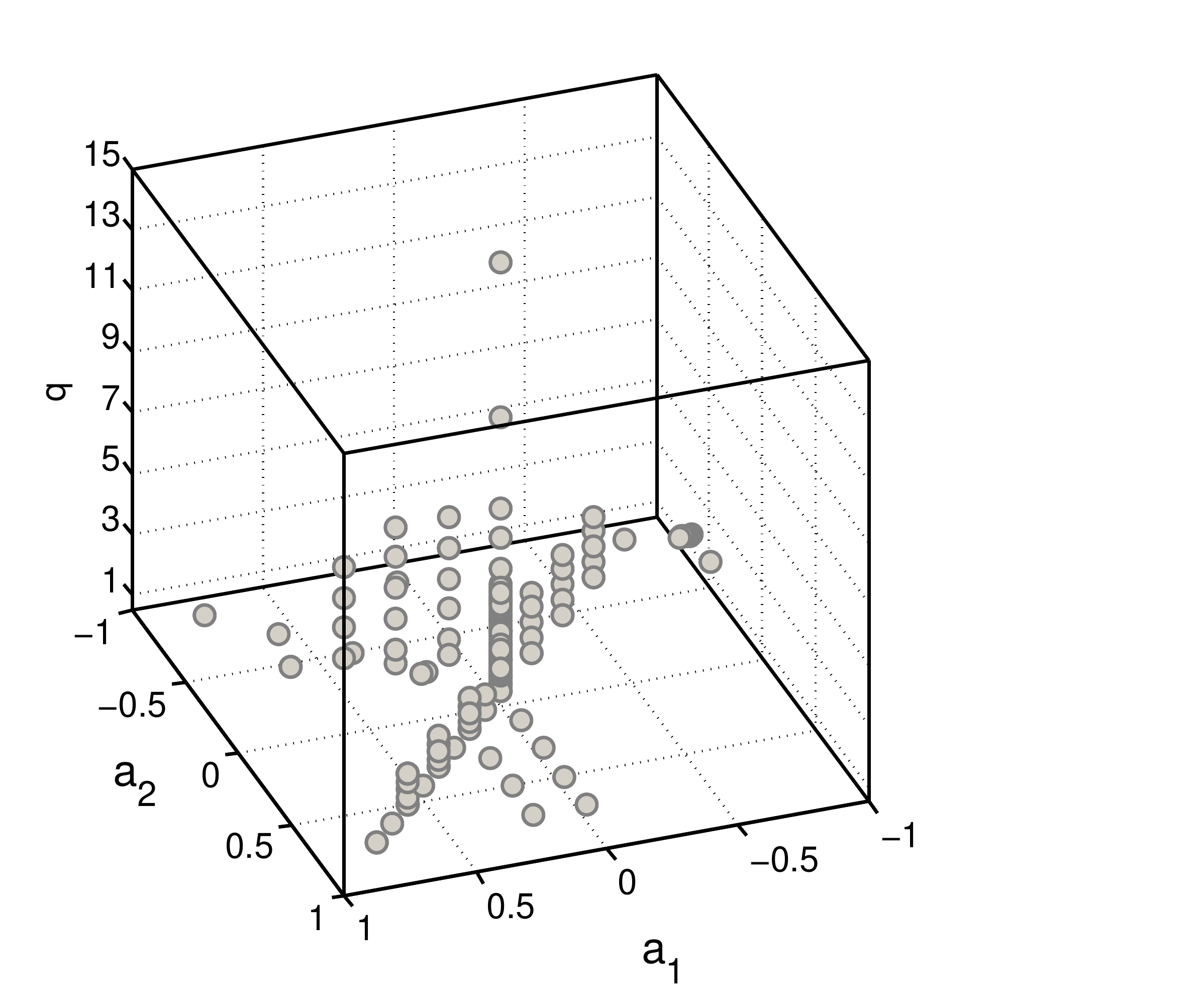}
\\
\includegraphics[scale=.3,trim= 450 100 100 100,clip]{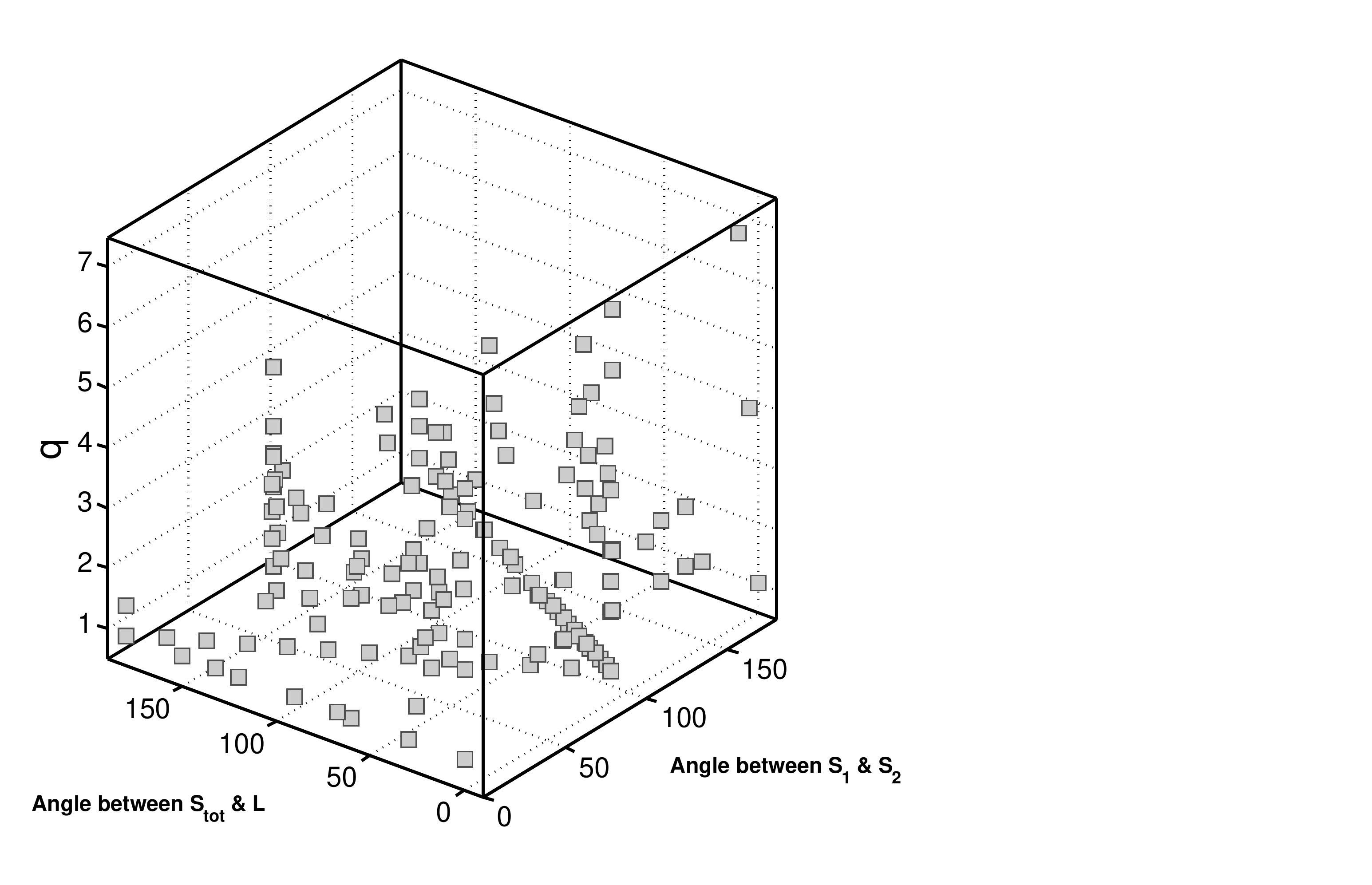}}
\caption{The left figure is GT's non-precessing runs and the right the precessing runs, The left plot indicates our set of runs, one for each circle, corresponding to the mass ratio, $q=m_1/m_2$ where $m_1$ and $m_2$ are the BH masses, and the initial spin of each black hole, $a_1=S_1/m_1^2$ and $a_2=S_2/m_2^2$.  The figure on the right, again depicts each run now by a square bot plots versus mass ratio on the verticle axis, $q$, and the angle between the total spins $S_{tot} = S_1+S_2$ and $L$, the total momentum and the angle between $S_1$ and $S_2$.}
\label{f:runs}       
\end{figure}

One of the challenges in NR is covering the full parameter space of the \bbh{} system.  There are no stringent constraints on the initial spins of the black hole, although there are some expectations~\citep{Gerosa:2013laa,Kesden:2009ds,Schnittman:2004vq}; and, in fact, we expects GWs to be the main method for observing  \bh{} parameters.   A similar scenario holds for the mass ratio, $q=m_1/m_2$ where $m_1$ and $m_2$ are the \bh{} masses, and total mass, $M=m_1+m_2$.  Recent work points to the possibility of higher mass \bh{s} than expected~\citep{Dominik:2014yma} and see the proceedings by Thomas Bulik in this collection.  Simulating extreme spins~\citep{Lovelace:2011nu} and mass ratios~\citep{Lousto:2010ut} is still a challenge, especially in producing quality waveforms from such simulations.   Despite these challenges, several groups are running generic, precessing \bbh{} systems.  The SXS collaboration has a published catalog of 174 BBH waveforms~\citep{Mroue:2013xna} including spins of 0.99 and mass-ratios of  1:10~\citep{Lovelace:pc},  RIT has generic runs~\citep{Lousto:2009ka} and GT~\citep{Pekowsky:2013ska} has several hundred shown here in fig.~\ref{f:runs}. 

The open problem and important challenge for NR and GW data analysis is in preparing to detect and characterize the parameters from a precessing \bbh{} system, see Ref~\citep{Hannam:2013pra} and references therein.

\section{Role of Higher Harmonics on BBH Mergers and GWs}
\label{modes}
\begin{figure}[h]
\hbox{
\includegraphics[scale=.35,trim= 50 170 50 200,clip]{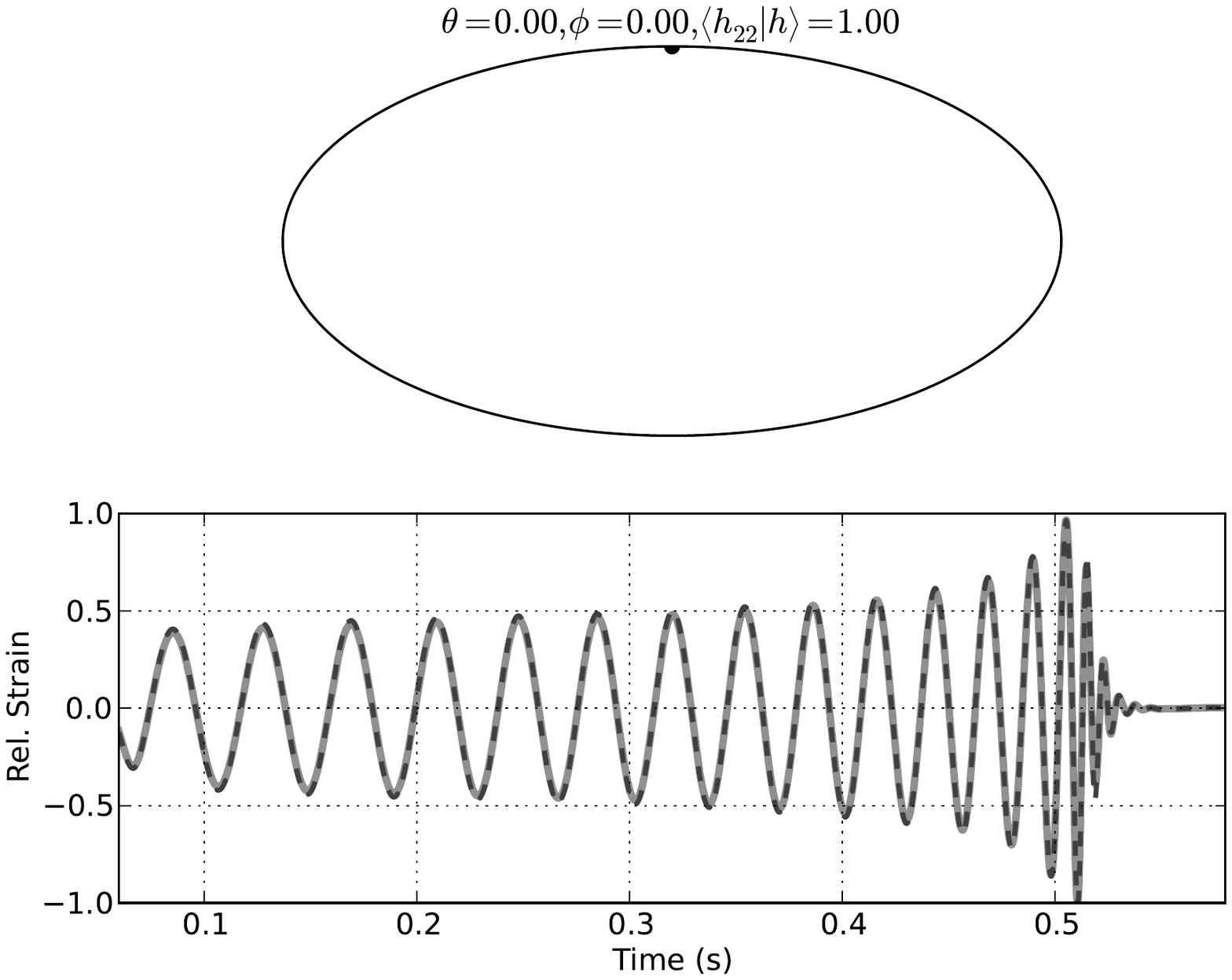}
\\
\includegraphics[scale=.35,trim= 50 170 50 200,clip]{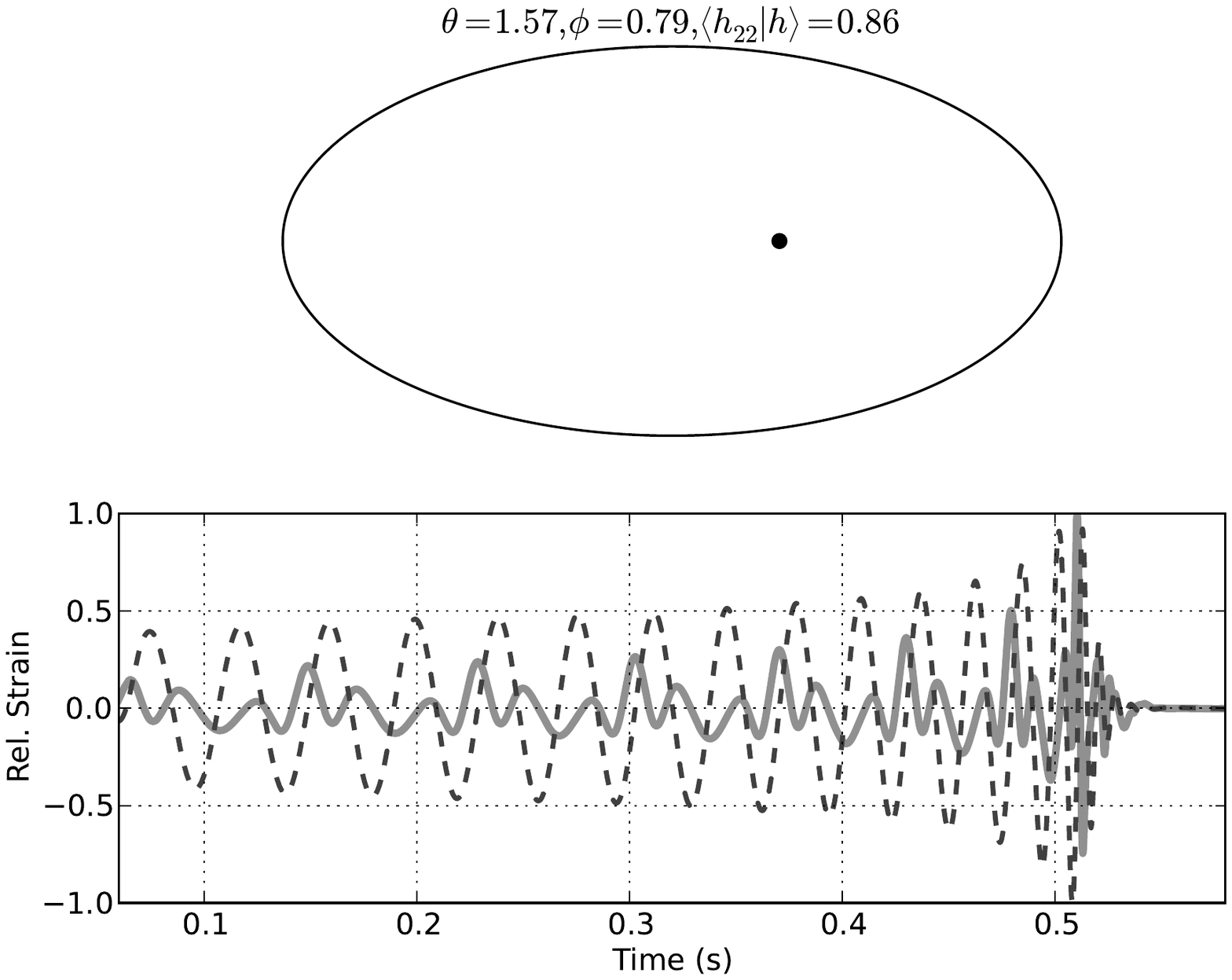}}
\caption{The upper figures show the location of the $q=10$ binay with respect to the detector.  The lower plots are of the strain of the $\ell=|m|=2$ waveform and the full waveform plotted on top of each other.  In the left, the binary is optimally oriented and the right, randomly oriented.  The strains are scaled to be one at the peak. }
\label{f:modes}       
\end{figure}

Gravitational waves encode the dynamics and history of the compact binary coalescence.   When the GW reaches the detector, the radiation also encodes the position of the binary on the sky and its inclination with respect to the detector(s).  In NR, we choose to represent the GWs extracted from our simulations via spin-weighted spherical harmonics that map the waveform from a function of angles to a function of modes, or harmonics, labeled by $\ell$ and $m$.  This mode description can be very useful theoretically since inspiralling binaries will have the quardupolar mode dominating, given by $\ell=|m|=2$.  However, as the black holes inspiral, merge and settle down to its final \bh{}, the information about the system is radiated in those harmonics.  In addition, the sky position also mixes in with these parameters as seen in fig.~\ref{f:modes} where we plot a \bbh{} system at two different lines of site with respect to the detector.  The binary in this figure has a mass ratio 1:10 and is non-spinning. To the left, you see that the binary was optimally oriented and the $\ell=|m|=2$ mode, given by $h_{22}$, overlaps with the full waveform, $h$, containing all the non-zero modes well. The overlap is the scalar product between $h$ and $h_{22}$ normalized to one.  To the right, we place the binary at a non-optimal position and now see that  $h_{22}$ no longer as good of a representation of the full waveform along that line of site.  

The question then is, for the current set of detectors, how important is this information?  Can we detect binaries of generic initial states without accounting for the modes?  If so, how does it bias the extracted parameters?  
This topic is addressed  in a recent set of papers~\citep{Healy:2013jza,Pekowsky:2012sr,Capano:2013raa} that are reviewed here.

\begin{figure}[h]
\hbox{
\includegraphics[scale=.35]{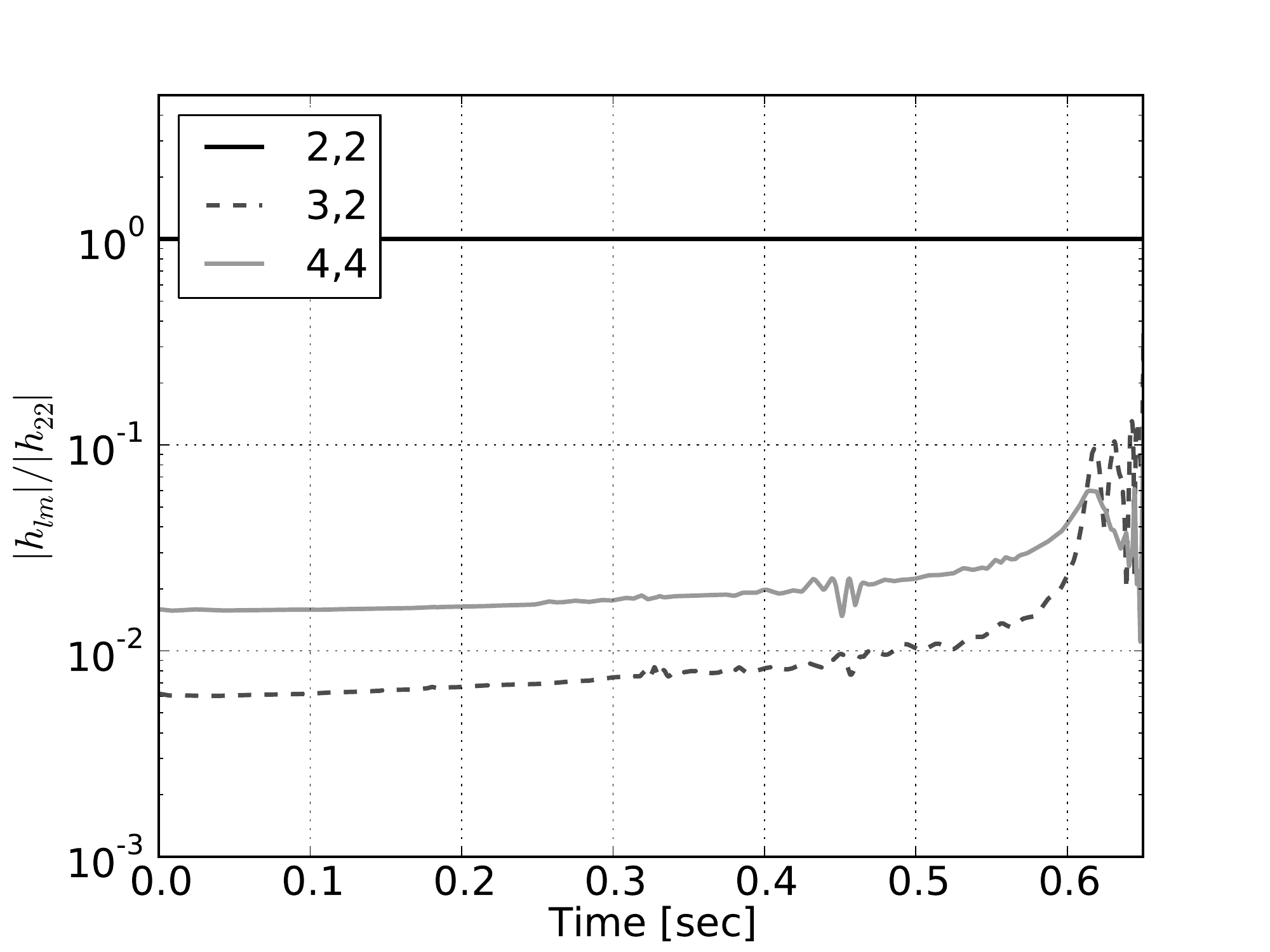}
\\
\includegraphics[scale=.35]{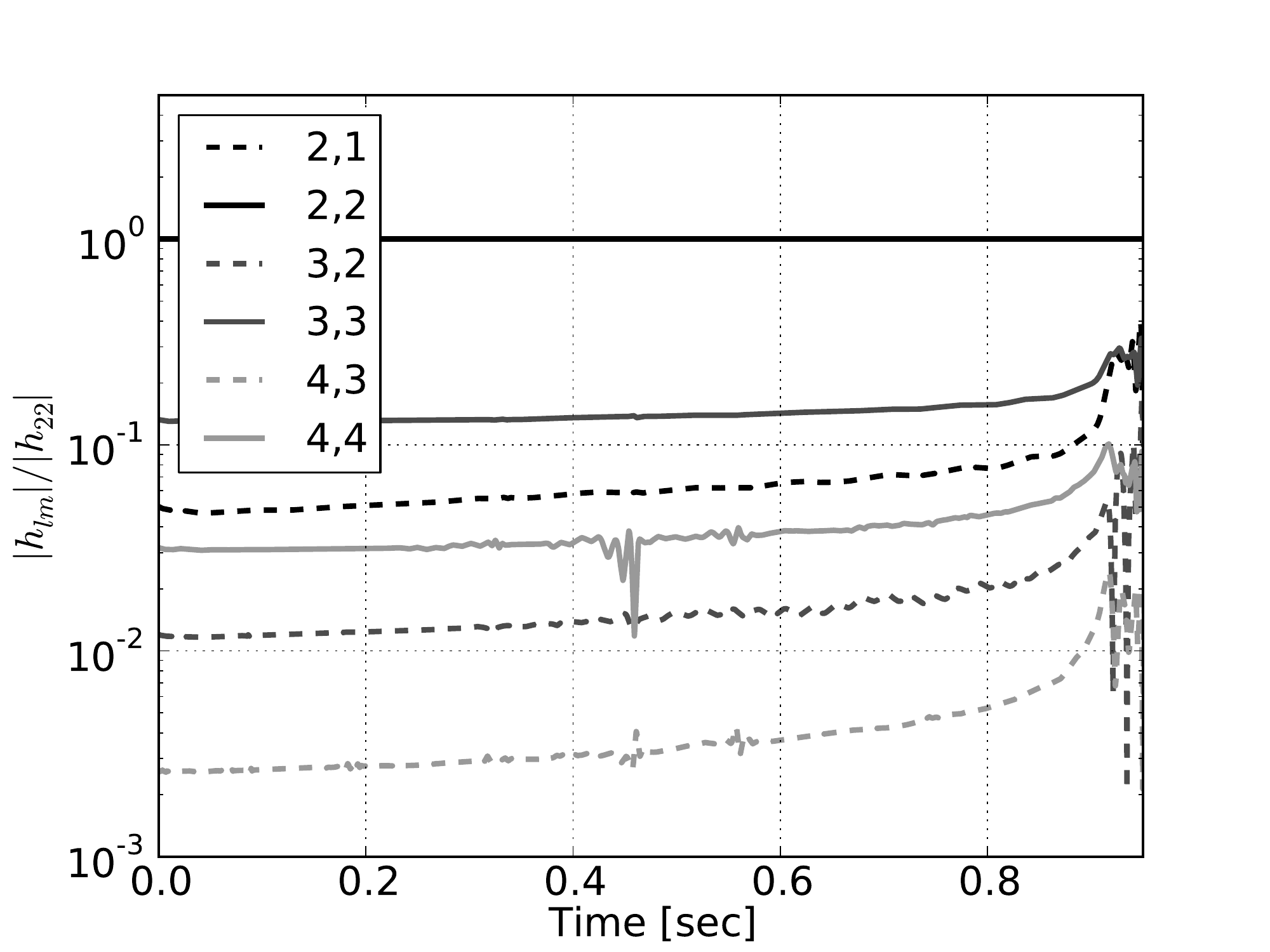}}
\caption{Figure from~\citep{Healy:2013jza} shows  the strain  amplitude  relative to the $\ell=|m|=2$ mode.  The left panel shows the case of an equal mass binary, and the right panel that of a 1:4 mass ratio system. In both figures the systems have been scaled to total mass $M = 100M_\odot$ and neither have spin. }
\label{f:ratios}       
\end{figure}

First we can look back at fig.~\ref{f:modes} and see which modes are needed to optimize our sky coverage.  We know that for highly precessing binaries or binaries with moderately large mass ratios, the energy in the radiation is spread out among the modes. In fig.~\ref{f:ratios} we see the relative amplitudes of higher modes in comparison with the $\ell=|m|=2$ mode for both an equal mass (left) and unequal (right). Clearly, as the mass ratio increases, the higher modes gain in amplitude.  In fact, the relative importance of the modes also increases towards merger, which makes these modes and hence NR potentially more important for BBH systems.
However, when we looked at  sky coverage, defined  as the percent of sky in a source-centric frame that a template will match with nature, which modes were important could deviate from a ranked list of amplitudes. To cover the entire sky to some measure, as mass ratio increases beyond $q=7$, more than 4 modes are necessary.  In the absence of precession, increasing spin magnitude reduces the number of needed modes.  Precessing systems required up to 12 modes for the cases studies in Ref.~\citep{Healy:2013jza}.

\begin{figure}[h]
\centerline{
\includegraphics[scale=.35,trim= 0 170 50 200,clip]{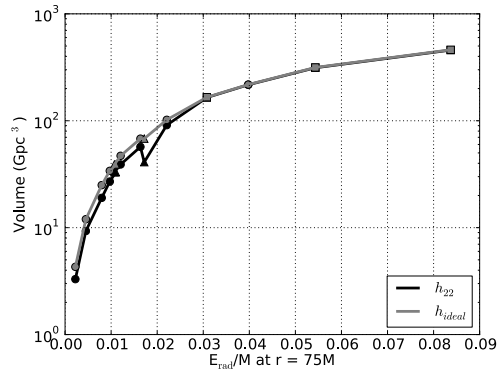}}
\caption{Figure from~\citep{Pekowsky:2012sr} shows the correlation between the total energy radiated and the detector reach measured in volume  using just the $\ell=|m|=2$ waveform or the full waveform labeled here as $h_{ideal}$. Circles label non-spinning  systems, squares are spinning but non-precessing systems, and triangles are precessing systems. }
\label{f:vol}       
\end{figure}

We then computed how the reach of a single detector would be impacted by not including all the relevant modes in Ref.~\cite{Pekowsky:2012sr}.  Figure~\ref{f:vol} is a plot from that paper that shows the energy radiated per run in both the $\ell=|m|=2$ only and all the modes versus the volume of the detector for that system.  The energy radiated in these systems corresponds to the amplitude of the waveform.  The actual vales in Gpc are not to be used literally since these numbers do not incorporate event-loss and other important aspects of the real detector pipeline.  We see that the fact the more modes are required for higher mass-ratio systems is diluted by the fact that these systems radiated less in total output; and, therefore, are reduced in volume.   Likewise, the higher modes are less important as the spin increases, seen here as the points on the right that radiate the largest amount of energy.  Precessing systems will depend on the parameters but the largest deviation between all the modes and just the quadrupolar is seen.

A further step was taken to understand the role of the sub-dominate modes for mass ratios of $q<4$ in ref~\citep{Capano:2013raa}.  They found that the higher modes did not improve detection rates for advanced LIGO when using a re-weighed SNR~\citep{Colaboration:2011np,Aasi:2012rja} and in fact including the modes increases the false-alarm rate. 
We may conclude, that for unequal-mass binaries of $q<4$, higher modes should not be included in the detection pipeline, although a definitive answer would require a all-mode injection into the real detector noise.  Whether or not that will remain the case for generic runs with precessing spins is an open question, as is the potential for parameter bias in the unequal-mass case which we are addressing in a future work~\citep{Pekowsky}.

\section{End State of BBH Merger}
\label{endstate}
\begin{figure}[h]
\centerline{
\includegraphics[scale=0.3]{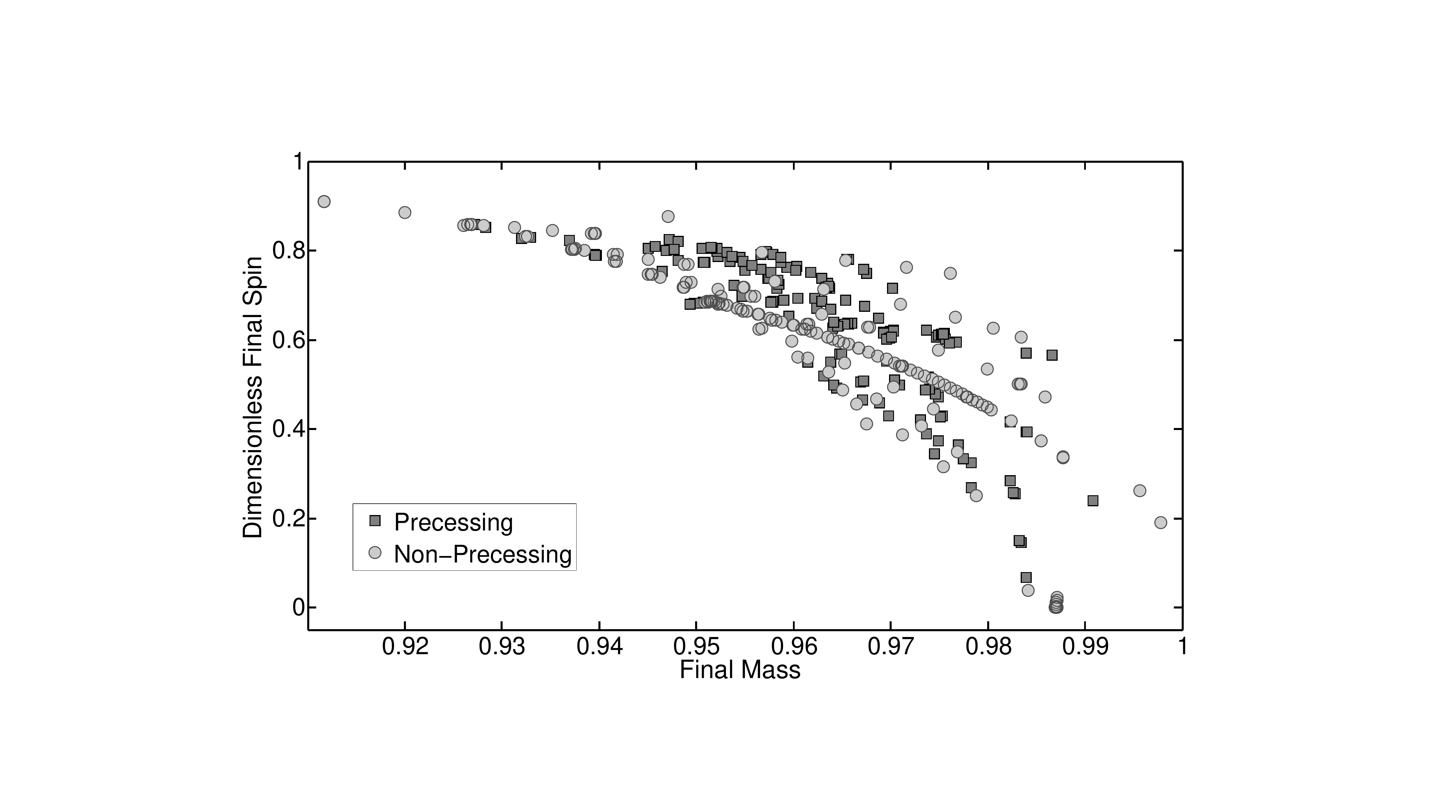}
}
\caption{Plotted are the spin and mass for the final, merged BH at the end of our NR simulations.  Each point represents a different simulation from unique initial data.  The squares are precessing runs and the circles are non-precessing.}
\label{f:final}       
\end{figure}

We've studied how the initial data of the binaries is imprinted on the harmonics of the waveforms.   In these final moments of the coalescence, the two \bh{s} merge into a perturbed Kerr \bh{}, whose gravitational radiation \textit{rings down} like a struck bell.  
With the \bh{'s} mass and spin known, perturbation theory provides  the \qnm{s} that describe ringdown \citep{PreTeu73_2,leaver85,PhysRevD76Berti}; however, perturbation theory cannot inform us of the amplitudes of each mode.  Recent studies have used NR waveforms to correlate the \qnm{} excitation amplitudes with initial parameters~\citep{PhysRevD76Berti,Kamaretsos:2011um,Kamaretsos:2012bs,Kelly:2012nd,London:2014cma}.  Figure~\ref{f:final} depicts the end state of our GT runs with the final BH's spin versus its mass.   \footnote{Note that the final mass is constrained to be less than one.   The mass and spin are computed using isolated horizons~\citep{Dreyer:2002mx} }

In ref~\citep{London:2014cma}, we studied the \qnm{s} excited by a series of NR simulations that focused on non-spinning, unequal mass binaries.  We found a rich variety of fundamental \qnm{s}, as well as overtones, and evidence for 2nd order  \qnm{s}.  Rather than the spin-weighted spherical harmonics used in NR extraction, we represent the waveforms in terms of the spheroidal harmonics which are natural to perturbation theory.   It has been an open question whether or not such a difference would make an impact on the modes. Inner-products between spherical and spheroidal harmonics suggest that for the $\ell = m$ modes, very little difference should be expected \citep{PhysRevD.73.024013}.  For the $\ell \neq m$ modes we expect ``beating'' that arises in the \qnm{s} when found using spherical harmonics which are combinations of the spheroidal harmonics used in perturbation theory \citep{Kelly:2012nd}.  As the final spin magnitude increasing deviates from 0, the difference between spherical and spheroidal harmonics is more pronounced, as expected \citep{London:2014cma}.

\begin{figure}[h]
\centerline{\hbox{
\includegraphics[scale=.25]{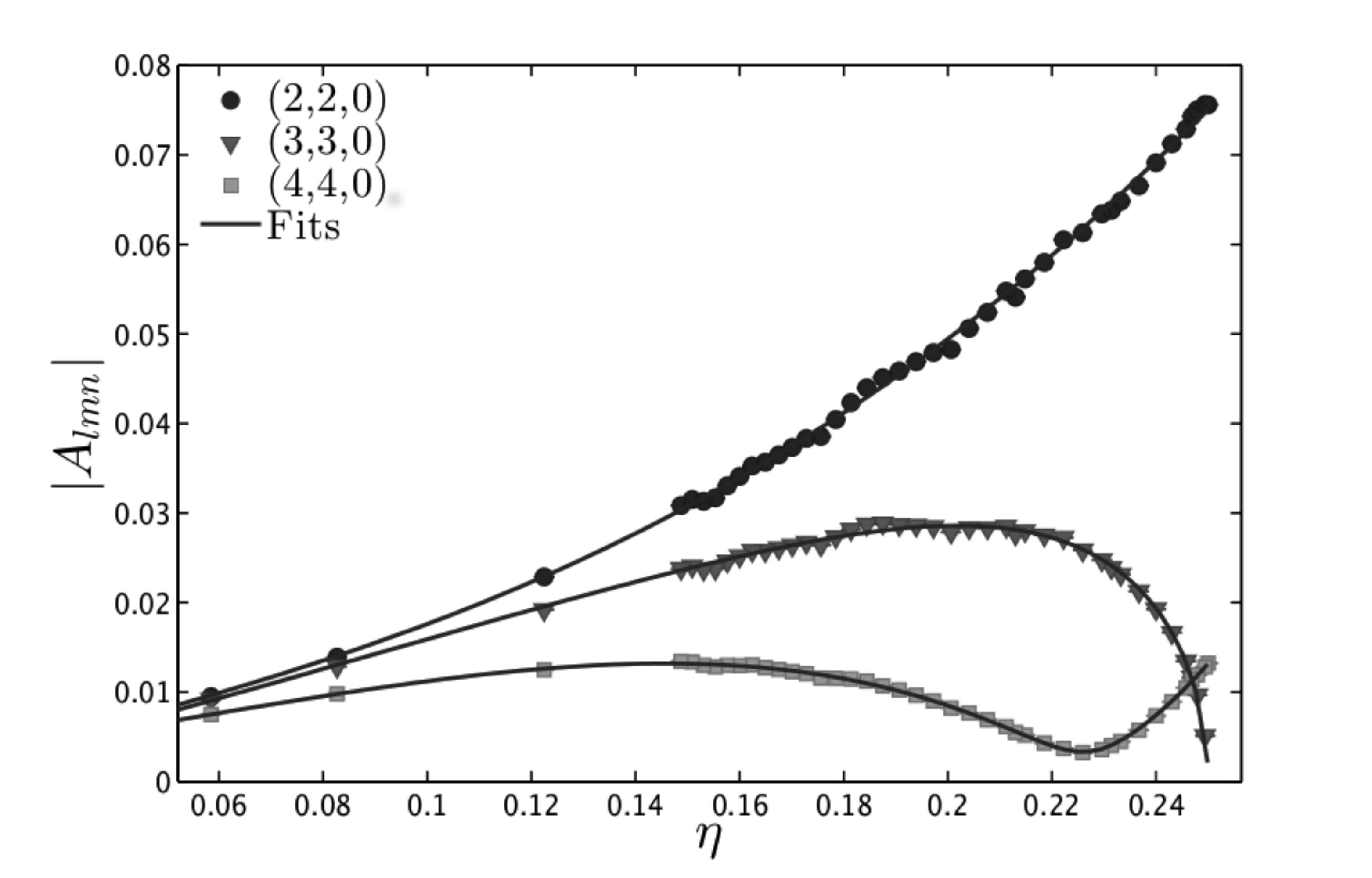}
\includegraphics[scale=.28,trim= 50 290 50 50,clip]{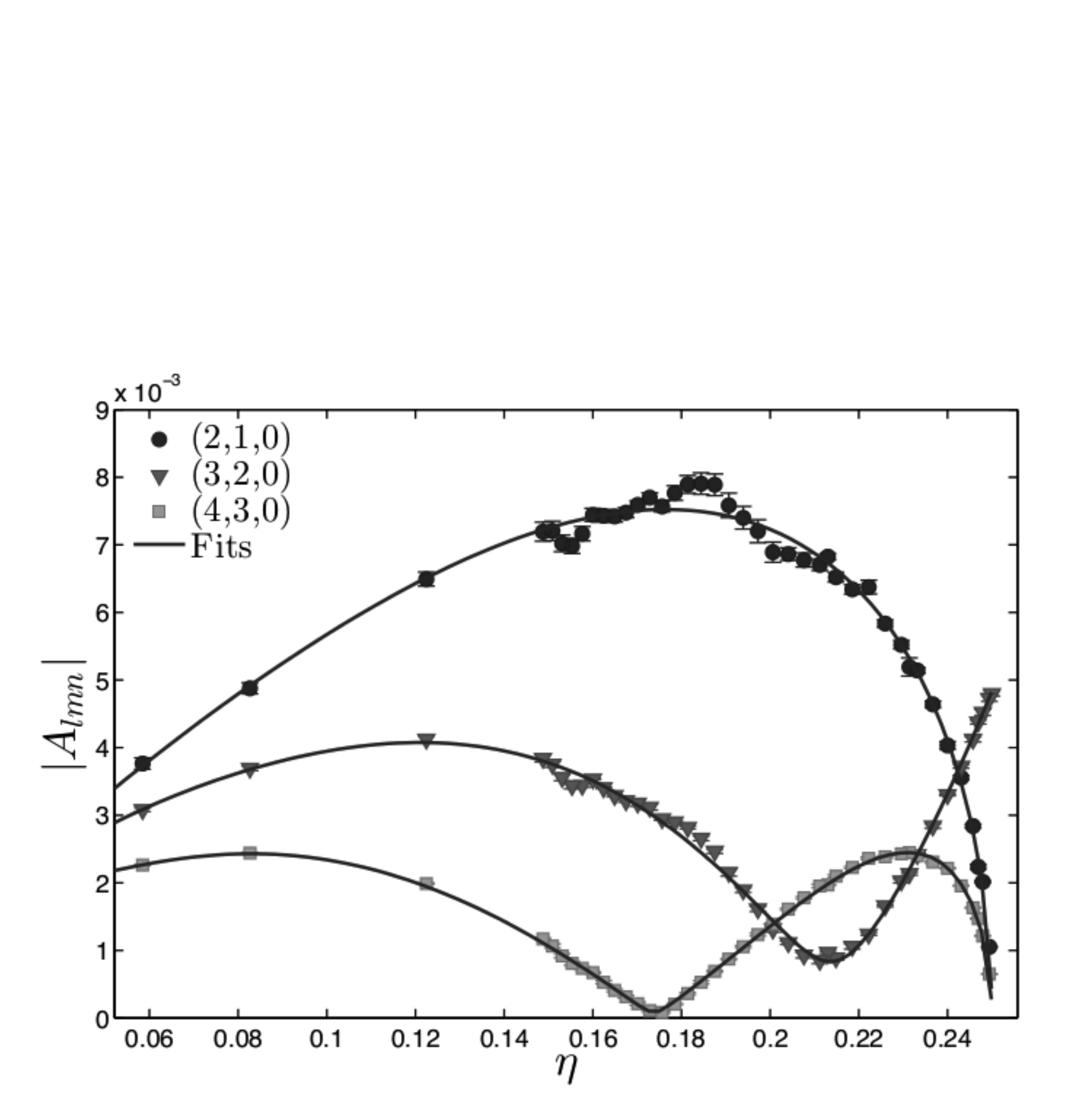}}}
\caption{Figure from~\citep{London:2014cma}  of the amplitudes of three \qnm{s} versus symmetric mass ratio and the corresponding fits.}
\label{f:lionel}       
\end{figure}

We find that the mass-ratio dependence of quasi-normal mode excitation is very well modeled by a Post-Newtonian like sum for most but not all \qnm{s}. We presented new fitting formulas for the amplitudes versus the mass ratio in Ref.~\citep{London:2014cma}.  Figure~\ref{f:lionel} plots the amplitudes of the fundamental \qnm{s} versus $q$ and include our fits.

\begin{figure}[h]
\centerline{
\includegraphics[scale=0.24]{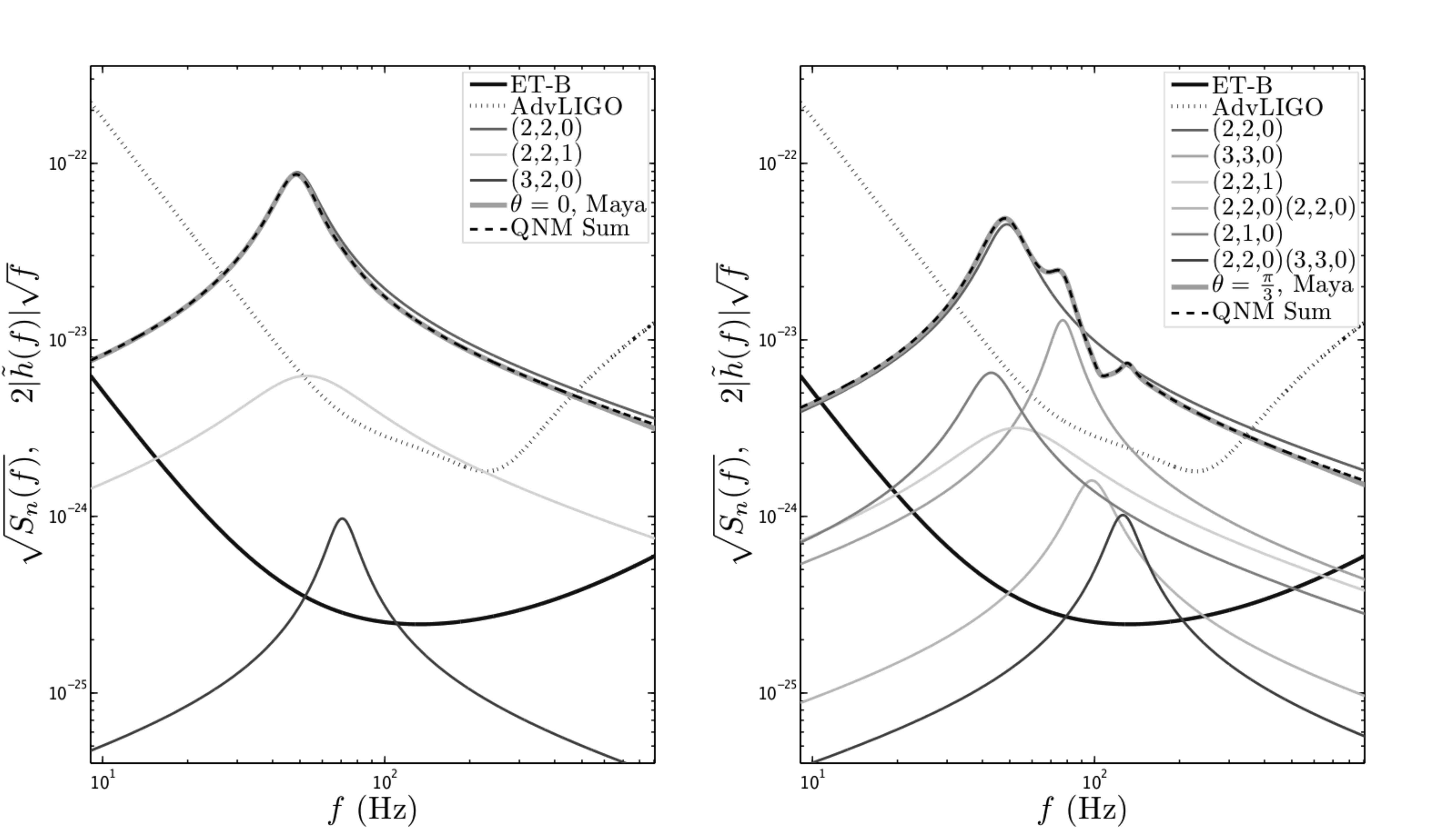}
}
\caption{This figure from~\citep{London:2014cma} plots the frequency domain envelopes of strain and fitted QNM amplitudes for a 2:1 mass ratio system of $350M_\odot$ at a distance of 100 Mpc. The left figure is for an optimally oriented binary and the left for a specific orientation.  Noise curves for the Einstein Telescope and Adv. LIGO are shown for reference. }
\label{f:snr}       
\end{figure}

The modes would be difficult to detect for current GW detectors.  However, there is the possibility that future detectors might be able to see them, as seen in fig.~\ref{f:snr} which depicts both Einstein Telescope and advanced LIGO noise predictions.  It seems possible to do gravitational-wave astronomy with \qnm{s}, if not as simply as first laid out in Refs.~\citep{Kamaretsos:2011um,Kamaretsos:2012bs}.  

\section{Mergers as Transient GW Events}
\label{bursts}
 
The LIGO and Virgo Collaborations search for gravitational wave transients, or {\em bursts}, without templates for sources like supernova or the unknown~\citep{Abadie:2012rq,Andersson:2013mrx}. Recently, several groups have begun to explore the merger of \bh{s} as a transient event~\citep{Fischetti:2010hx,Clark,klimenko,cornish}.  This is a very interesting development that focuses on intermediate-mass \bbh{} systems that may only have a few cycles in band.  See \citep{Clark} in this volume for a discussion of one such approach to mergers as transients.

\section{Conclusions}
\label{concl}

Undoubtably, numerical relativity, analytical relativity and gravitational wave data analysis have made great strides in the last few years.  With the advanced LIGO scheduled to begin taking data next year and her sister detectors soon after, the theoretical efforts are focusing down on the crucial open questions for detecting GW sources and estimating their parameters.  This proceeding highlighted just a few of those questions.   

Clearly, the parameter space of generic \bh{} binaries is large and the NR community continues its efforts to provide waveforms for the full space and to turn these NR waveforms into viable GW templates.   One avenue of research that is still open is using tools like principal component analysis~\citep{Clark} to predict where we need a new NR run that will make the most impact on the template bank.  

Over the last couple of years, work has been done to measure the impact of higher modes on detection.  We are optimistic that for lower mass-ratios and non-precessing spins, the dominate mode will be enough for detection. The problem is still open for a verification of this statement as mass ratio increases over the full \bbh{} total mass range.  As is the question of higher modes for highly precessing binaries where the $\ell=|m|=2$ mode is not always dominate.  

NR runs result in the \bbh{} system ringing down to the final \bh{}.  We have found a rich and fascinating regime during this epoch, revealing a variety of \qnm{s}.  While it is still questionable whether these modes will be detectable by the current generation of GW detectors, it does hint at the tantalizing possibility of conduction GW astronomy with \bh{} ringdown waveforms.

NR still has a story to tell GW astronomy.  New and exciting work is being done in \bh{}-neutron star and binary neutron stars.  \bbh{s} as transients are being seriously investigated.   At the end of the day, it will be \gw{} astronomy that will excite new work in NR as we learn about how general relativity is manifested in the Universe.  

\begin{acknowledgement}
The authors thank the organizers and participants of the Sant Cugat Forum on Astrophysics on Gravitational Wave Astrophysics.  The work presented here was supported by NSF PHY-0955825.
\end{acknowledgement}

\bibliographystyle{apalike}      
\bibliography{biblio}


%
%

\end{document}

%% file: localdefs.tex
\def\gw#1{gravitational wave#1 (GW#1)\gdef\gw{GW}}
\def\nr#1{numerical relativity#1 (NR#1)\gdef\nr{NR}}

\def\imbh#1{intermediate mass black hole#1(IMBH#1)\gdef\imbh{IMBH}}
\def\smbh#1{supermassive black hole#1(SMBH#1)\gdef\smbh{SMBH}}
\def\bbh#1{binary black hole#1 (BBH#1)\gdef\bbh{BBH}}
\def\bh#1{black hole#1 (BH#1)\gdef\bh{BH}}
\def\ns#1{neutron star#1 (NS#1)\gdef\ns{NS}}
\def\gw#1{gravitational wave#1 (GW#1)\gdef\gw{GW}}
\def\pnw#1{post-Newtonian#1 (PN#1)\gdef\pnw{PN}}
\def\eos#1{equation of state#1 (EOS#1)\gdef\eos{EOS}}
\def\grb#1{gamma-ray burst#1 (GRB#1)\gdef\grb{GRB}}
\def\amr#1{adaptive mesh refinement#1 (AMR#1)\gdef\amr{AMR}}
\def\isco#1{innermost stable circular orbit#1 (ISCO#1)\gdef\isco{ISCO}}
\def\qnm#1{quasi-normal modes#1 (QNM#1)\gdef\qnm{QNM}}